\documentclass[journal]{IEEEtran}
\usepackage[T1]{fontenc}
\usepackage[utf8]{inputenc}
\usepackage{amsmath,amssymb}
\usepackage{graphicx}
\usepackage{caption}
\usepackage{subcaption}
\usepackage{tikz}
\usepackage{tikz-3dplot}
\usetikzlibrary{calc,arrows.meta}
\usepackage{pgf}            % needed by cairolatex gnuplot output
\usepackage{booktabs}       % \toprule / \midrule / \bottomrule
\usepackage{multirow}       % \multirow in tables
\usepackage{listings}
\usepackage{xcolor}
\usepackage{hyperref}
\usepackage{cite}
\usepackage{float}
\usepackage{todonotes}

\title{Stencil Computation at the Intersection of AI and HPC}

\author{Timoth\'ee~Ewart and Mauricio~Araya-Polo
\thanks{T. Ewart is with Intel Corporation, Santa Clara, USA}
\thanks{M. Araya is with TotalEnergies EP Research and Technology US, Houston, USA}}
\begin{document}

\maketitle

% ---------------------------------------------------------------
\begin{abstract}
Tensor compilers such as TinyTC and OpenAI Triton were originally
developed for AI workloads, but the same tiling and memory abstractions
can be applied to implement efficient high-order stencils for scientific
and industrial applications.
We demonstrate this for an 8th-order, 25-point acoustic stencil with boundary conditions over an a demanding-sized grid, targeting GPGPUs, where we compare the hardware-specialized TinyTC
implementation with a portable PyTorch/Triton implementation.
The target platforms for evaluation include Intel B70, B580, GPU~MAX~1550,
NVIDIA A100/RTX6000 Blackwell/H100, and AMD MI325x.
For instance, on Battlemage B580 TinyTC reaches 15.6~Gpts/s versus 13.5~Gpts/s for
PT/Triton under random initialization, while zero-initialized runs reach
up to 35.8~Gpts/s due to hardware memory compression.
Using roofline and memory-hierarchy profiling, we show that -as expected- performance
is predominantly bandwidth-limited and that compiler-managed L1/LSC
caching can effectively replace programmer-managed shared-memory staging
for this stencil class.
Overall, the results position TinyTC as the performance-oriented path on
Intel hardware and PyTorch/Triton as a strong portability/productivity
baseline for cross-vendor HPC stencil development.
\end{abstract}


% ---------------------------------------------------------------
\section{Introduction}
\label{sec:intro}

Stencil computation is one of the oldest and most enduring patterns
in scientific computing.
At its core, a stencil kernel updates each point of a grid by
combining values from its nearest neighbors, making it the
foundation of finite-difference methods for solving partial
differential equations (PDEs)~\cite{LeVeque2007}.
Applications range from seismic wave propagation~\cite{AkiRichards2002} and 
fluid dynamics~\cite{FerzigerPeric2002} to image processing~\cite{GonzalezWoods2017}.
Despite its conceptual simplicity, achieving high performance for
stencils has remained a moving target for over two decades,
constantly reshaped by the evolution of computing hardware (CPU, GPU and now NPU/TPU),
and software paradigms and programming models (OpenMP, CUDA, HIP, Tile Compiler, etc.). We retrace this journey
in the next paragraphs to contextualize our work and highlight the key developments.

\paragraph{Multi-core era (2000s)}
In the early 2000s, the shift from single-core to multi-core
processors \cite{IntelXeonDualCore2006} forced stencil developers to adopt shared-memory
parallelism.  Cache-blocking and loop tiling became essential
optimizations to exploit spatial and temporal locality across
multiple cores~\cite{Wonnacott2000,datta2008}.
Auto-tuning research~\cite{Datta:EECS-2009-177}
demonstrated that algorithmic parameters could be searched
automatically for optimal performance rather than hand-tuned.

\paragraph{Heterogeneous accelerators (2005--2013)}
The IBM Cell Broadband Engine (PowerXCell in the HPC variant) was the first mainstream accelerator, (although dedicated accelerators had already existed in the 1980s, such as the MicroVAX matrix accelerator) Introduced in 2005 for the
PlayStation~3 and the Roadrunner supercomputer, introduced a
radically different model: one general-purpose (PowerPC) core orchestrating
eight lean SIMD (128-bit width) engines (SPEs) over a local store (256 KB) with explicit
DMA transfers~\cite{cell_stencil,Williams2006Cell}. It was the dawn of accelerator for HPC. Stencil codes had to be restructured around explicit data movement \cite{delacruz2014semistencil} and software-managed memory, foreshadowing the challenges that
would later arise on GPU architectures.
This period established that stencil performance is driven by memory bandwidth, 
not compute, a fact reflected in roofline analysis~\cite{roofline}.

\paragraph{GPU era (2010s--present)}
General-purpose GPU computing (GPGPU), enabled by CUDA (released 2004-2007) and later
OpenCL, brought massive thread-level parallelism to stencil
workloads alongside large global memory (up to several GB) directly accessible by threads, unlike the Cell's mandatory local store management. GPUs offer shared memory,
a much larger number of registers - 64k per Streaming Processor (H100), and warp-level primitives \cite{LinGrover2018Warp} as optional but powerful
latency-hiding optimizations~\cite{datta2008,micikevicius2009,GrosserCohenKellyRamanujamSadayappanVerdoolaege2013,HolewinskiPouchetSadayappan2012}.
The period produced a rich body of work on stencil
kernels for NVIDIA GPUs, and later on directive-based
portability layers (OpenACC, OpenMP target) that could target both
CPU and GPU backends \cite{ShanArayaPolo2024StencilGPU}.

\paragraph{AI accelerates HPC (2020--present)}
The explosion of deep learning has driven unprecedented investment
in GPU hardware and in programming abstractions for tensor
operations: CUTLASS \cite{NVIDIA_CUTLASS}, XeTLA \cite{Intel_XeTLA}, and tile compilers such as
TinyTC~\cite{tinytc} for Intel or CuTile~\cite{cutile} for NVIDIA,
lower the level of abstraction just enough
to expose the hardware tiling and register-file hierarchy without
requiring hand-written assembly.

In parallel, OpenAI Triton~\cite{triton} has emerged as the
main de facto tile programming model, offering a
tile-based Python DSL with a low barrier to entry; it has
been widely adopted in both industry and research.
Originally designed for fused transformer kernels, Triton is
expressive enough to describe any regular tiled computation and
supports multiple GPU backends out of the box. Tensor compilers target primarily AI workloads: matrix multiplications,
attention, convolutions, yet stencil computations share the
same tiling structure and memory access patterns that make them
attractive candidates for evaluation.

Our contribution is to
%Our primary target is the Intel Battlemage~580 GPU, on which we
evaluate a TinyTC implementation as the architecture-specialized path
and PT/Triton as the portability baseline.
Because the stencils are bandwidth-limited, we focus on data reuse and
cache behavior, with particular attention to L1/LSC efficiency.
A preliminary version of this work was presented as a poster at ISC 2026~\cite{ewart2026iscposter}.

In more concrete terms, the paper makes three linked contributions: \begin{enumerate}
\item a comparative evaluation of
TinyTC and PT/Triton for a 25-point acoustic stencil
with boundary conditions; 
\item a memory-centric analysis (throughput, roofline, and
cache/DRAM counters) that explains the observed performance trends,
including the impact of the B580 memory compression; and 
\item a
cross-vendor PT / Triton comparison with different GPUs to characterize portability/performance trade-off.
\end{enumerate}

In general, our goal is to answer whether a specialized tile compiler and a
mainstream AI programming model can serve as first-class tools
for HPC stencil development.

The remainder of this paper is organized as follows.
Section~\ref{sec:related} reviews related work.
Section~\ref{sec:background} introduces the hardware platforms and
programming models.
Section~\ref{sec:impl} describes the stencil implementations.
Section~\ref{sec:setup} presents the experimental setup.
Section~\ref{sec:results} reports and discusses performance results.
Section~\ref{sec:conclusion} concludes.

% ---------------------------------------------------------------
\section{Related Work}
\label{sec:related}

Stencil optimization on parallel architectures has been studied
extensively over the past two decades.
We briefly survey the main algorithmic strategies and position our
work with respect to the most closely related efforts.

\paragraph{Tiling, caching, and time skewing on CPUs}
The literature on this topic is vast; we highlight a representative subset.
\cite{FrigoLeisersonProkopRamachandran1999},
\cite{FrigoStrumpen2005},
\cite{FrigoStrumpen2006},
\cite{StrzodkaShaheenPajakSeidel2010},
\cite{TangChowdhuryKuszmaulLukLeiserson2011},
\cite{Wonnacott2000},
\cite{Wonnacott2002},
\cite{JinMellorCrummeyFowler2001},
\cite{McCalpinWonnacott1998},
\cite{SongLi1999}.
These papers cover cache-aware, cache-oblivious, and time-skewing
techniques to reduce memory traffic and exploit locality, since
bandwidth is the bottleneck and caching is key to performance.

For example, cache-blocking and time skewing~\cite{Wonnacott2000,FrigoLeisersonProkopRamachandran1999} were
among the first systematic techniques to improve data reuse for
stencil codes on multi-core CPUs.
Overlapped tiling~\cite{strzodka2011} extended this idea to parallel
execution by introducing redundant computation at tile boundaries in
exchange for reduced memory traffic.
Split tiling~\cite{split_tiling} avoids redundant computation by
decomposing the time-skewed domain into two phases.

\paragraph{GPU-specific strategies and shared memory management}
The GPU literature is also very rich:
\cite{HolewinskiPouchetSadayappan2012}, \cite{nguyen2010}, \cite{micikevicius2009}, \cite{GrosserCohenKellyRamanujamSadayappanVerdoolaege2013}, \cite{an5d}, \cite{sai2020}.

For example, Micikevicius~\cite{micikevicius2009} proposed a simple but highly
effective 2.5D approach: tile a 2D plane into shared memory while
keeping the values along the third (streaming) dimension in
registers, effectively hiding global memory latency through the
register file. This approach be the starting point of our implementations.
Nguyen~et~al.~\cite{nguyen2010} introduced 3.5D
blocking, combining 2D spatial blocking in shared memory with 1D
temporal blocking to increase arithmetic intensity.
This strategy is particularly efficient for high-order stencils
because it avoids the shared-memory pressure that grows with stencil
order, and maps naturally onto the warp execution model.
The AN5D framework~\cite{an5d} further refines 2.5D and 3.5D
blocking with fixed register allocations, double buffering, and
division of the streaming dimension, achieving near-roofline
performance for simple single-statement kernels.

Sai~et~al.~\cite{sai2020} conducted a thorough empirical study of
a 25-point 8th-order acoustic isotropic stencil with PML boundary
conditions on multiple NVIDIA GPU generations (P100, V100, A100).
They systematically compare 3D and 2.5D blocking strategies,
shared-memory vs.~register-file usage. Several of their implementations
were shown to achieve twice the
performance of a proprietary OpenACC code.

Across these GPU approaches, performance hinges on effective use of the GPU
memory hierarchy. Explicit management of shared
memory (called Shared Local Memory, SLM, on Intel GPUs) remains a
central and labor-intensive concern: the programmer must manually
orchestrate data staging, synchronization barriers, and tile
boundaries to keep the fast scratchpad occupied, and in the most
hand-tuned variants even manage register usage explicitly.

A key motivation for tile compilers is to relieve the programmer of
this burden entirely.
Rather than relying on programmer-managed SLM. For instance, TinyTC targets the
\emph{L1 data cache} (Load Store Cache, LSC) of Intel GPU
architectures.
On Intel Battlemage and Intel GPU MAX1550, the LSC (384 KB) is significantly
larger than the SLM (64 KB), providing more capacity for data reuse without
requiring explicit staging.
The tile compiler generates the necessary access patterns and
prefetch schedules automatically, letting the hardware cache
hierarchy absorb the data reuse that would otherwise require
hand-written shared-memory code.
This shift: from programmer-managed scratchpad to
compiler-managed L1 (although TTC could do it), is one of the central design differences
between the TinyTC approach and classical GPU stencil
implementations, and is a key value proposition this paper evaluates.

\paragraph{Cerebras WSE-2 Integration} 
At the other end of the architectural spectrum,
 Jacquelin~et~al.~\cite{cerebras} evaluate the 25-point stencil onto the Cerebras WSE-2, a wafer-scale engine with ~850,000 cores, each with 48~KB of local SRAM and interconnected via a high-speed on-chip fabric. The X and Y dimensions of the grid are mapped onto the fabric, with each core exchanging data with its 16 neighbors using localized broadcast patterns, while the Z dimension resides entirely in the local memory of each PE. Incoming neighbor data is multiplied by stencil coefficients and accumulated in a local buffer. This careful orchestration of communication and computation makes the stencil effectively compute-bound (~503 TFLOPs), fully leveraging the WSE-2’s massive parallelism and on-chip memory.

\paragraph{DSL and compiler approaches}
Domain-specific languages such as Devito~\cite{devito} and
Halide~\cite{halide} generate optimized stencil code from
high-level specifications, while auto-tuning
frameworks~\cite{datta2008} explore the space of blocking
parameters at runtime.
At the SIMD CPU level, Yount~\cite{vector_folding} proposed
\emph{vector folding}, storing multi-dimensional data blocks
in SIMD registers to reduce memory traffic by up to~2.7$\times$
on Intel Xeon Phi.

\paragraph{Tile compilers for scientific stencils}
To our knowledge, \cite{wassell2025optimized} lays the first foundations for tile architecture in HPC, particularly for matrix–vector BLAS operations associated with sparsity-aware tile compression. Our work follows this direction, but we focus on tile compilers such as TinyTC and Triton on high-order scientific stencils with boundary conditions.

%Thiscomparing TinyTC as the architecture-aware implementation
%for Intel Battlemage 580 with PT/Triton as the mainstream portability baseline, both
%grounded in the Micikevicius 2.5D register-tiling strategy~\cite{micikevicius2009}.

% ---------------------------------------------------------------
\section{Background}
\label{sec:background}

\subsection{Stencil Computation}

We study a stencil-based solver for the acoustic isotropic
approximation of the 3-D wave equation, widely used in seismic
depth imaging by the oil and gas industry~\cite{sai2020}.
The Minimod proxy-application, from which the base code is extracted is developed
by TotalEnergies, it represents production-level geophysical applications and it solves the above mentioned equation using
high-order finite differences \cite{meng2020minimod}. 
%Its stencil radius is 4, this 25-point stencil solver
%exchanges nearest-neighbor halos at every time step.
The physical domain is a cubic grid of up to thousands points in every direction.
To solve the equation for realistic scenarios simulations iterate over a large number of time steps.
%to model the propagation of a seismic wavefield from a source
%perturbation.

\begin{figure}[t]
  \centering
  \tdplotsetmaincoords{65}{130}
  \begin{tikzpicture}[tdplot_main_coords, scale=1.15,
    every node/.style={font=\scriptsize}]
    % Coordinate macros
    \pgfmathsetmacro{\S}{4.0}      % outer cube side (= 800 cells)
    \pgfmathsetmacro{\p}{0.60}     % PML thickness, exaggerated (= 27 cells)
    \pgfmathsetmacro{\lo}{\p}      % inner cube low bound
    \pgfmathsetmacro{\hi}{\S-\p}   % inner cube high bound

    % ----------------------------------------------------------------
    % Inner cube hidden faces (for depth cue)
    % ----------------------------------------------------------------
    \fill[blue!15,opacity=0.75]
      (\lo,\lo,\lo)--(\hi,\lo,\lo)--(\hi,\hi,\lo)--(\lo,\hi,\lo)--cycle; % bottom
    \fill[blue!12,opacity=0.75]
      (\lo,\hi,\lo)--(\hi,\hi,\lo)--(\hi,\hi,\hi)--(\lo,\hi,\hi)--cycle; % back
    \fill[blue!18,opacity=0.75]
      (\lo,\lo,\lo)--(\lo,\hi,\lo)--(\lo,\hi,\hi)--(\lo,\lo,\hi)--cycle; % left

    % ----------------------------------------------------------------
    % Inner cube visible faces
    % ----------------------------------------------------------------
    \fill[blue!30,opacity=0.90]
      (\lo,\lo,\lo)--(\hi,\lo,\lo)--(\hi,\lo,\hi)--(\lo,\lo,\hi)--cycle; % front
    \fill[blue!25,opacity=0.90]
      (\hi,\lo,\lo)--(\hi,\hi,\lo)--(\hi,\hi,\hi)--(\hi,\lo,\hi)--cycle; % right
    \fill[blue!38,opacity=0.90]
      (\lo,\lo,\hi)--(\hi,\lo,\hi)--(\hi,\hi,\hi)--(\lo,\hi,\hi)--cycle; % top

    % ----------------------------------------------------------------
    % Inner cube edges (12 edges)
    % ----------------------------------------------------------------
    \draw[blue!70!black,thick](\lo,\lo,\lo)--(\hi,\lo,\lo);
    \draw[blue!70!black,thick](\hi,\lo,\lo)--(\hi,\hi,\lo);
    \draw[blue!70!black,thick](\hi,\hi,\lo)--(\lo,\hi,\lo);
    \draw[blue!70!black,thick](\lo,\hi,\lo)--(\lo,\lo,\lo);
    \draw[blue!70!black,thick](\lo,\lo,\hi)--(\hi,\lo,\hi);
    \draw[blue!70!black,thick](\hi,\lo,\hi)--(\hi,\hi,\hi);
    \draw[blue!70!black,thick](\hi,\hi,\hi)--(\lo,\hi,\hi);
    \draw[blue!70!black,thick](\lo,\hi,\hi)--(\lo,\lo,\hi);
    \draw[blue!70!black,thick](\lo,\lo,\lo)--(\lo,\lo,\hi);
    \draw[blue!70!black,thick](\hi,\lo,\lo)--(\hi,\lo,\hi);
    \draw[blue!70!black,thick](\hi,\hi,\lo)--(\hi,\hi,\hi);
    \draw[blue!70!black,thick](\lo,\hi,\lo)--(\lo,\hi,\hi);

    % ----------------------------------------------------------------
    % Outer cube edges (12 edges)
    % ----------------------------------------------------------------
    \draw[black,thick](0,0,0)--(\S,0,0);
    \draw[black,thick](\S,0,0)--(\S,\S,0);
    \draw[black,thick](\S,\S,0)--(0,\S,0);
    \draw[black,thick](0,\S,0)--(0,0,0);
    \draw[black,thick](0,0,\S)--(\S,0,\S);
    \draw[black,thick](\S,0,\S)--(\S,\S,\S);
    \draw[black,thick](\S,\S,\S)--(0,\S,\S);
    \draw[black,thick](0,\S,\S)--(0,0,\S);
    \draw[black,thick](0,0,0)--(0,0,\S);
    \draw[black,thick](\S,0,0)--(\S,0,\S);
    \draw[black,thick](\S,\S,0)--(\S,\S,\S);
    \draw[black,thick](0,\S,0)--(0,\S,\S);

    % ----------------------------------------------------------------
    % Dimension annotations
    % ----------------------------------------------------------------
    % Outer cube dimension (left-back vertical edge): 27 - 746 - 27
    \pgfmathsetmacro{\xoff}{-0.35}  % offset to the left of the edge
    \draw[<->,red,thick] (\xoff,\S,0) -- (\xoff,\S,\lo) 
      node[midway,right,black] {\small 27};
    \draw[<->,red,thick] (\xoff,\S,\lo) -- (\xoff,\S,\hi) 
      node[midway,right,black] {\small 746};
    \draw[<->,red,thick] (\xoff,\S,\hi) -- (\xoff,\S,\S) 
      node[midway,right,black] {\small 27};

  \end{tikzpicture}
  \caption{Domain decomposition for an $800^3$ grid showing the inner
    $746^3$ region (blue) within the outer domain.}
  \label{fig:domain}
\end{figure}
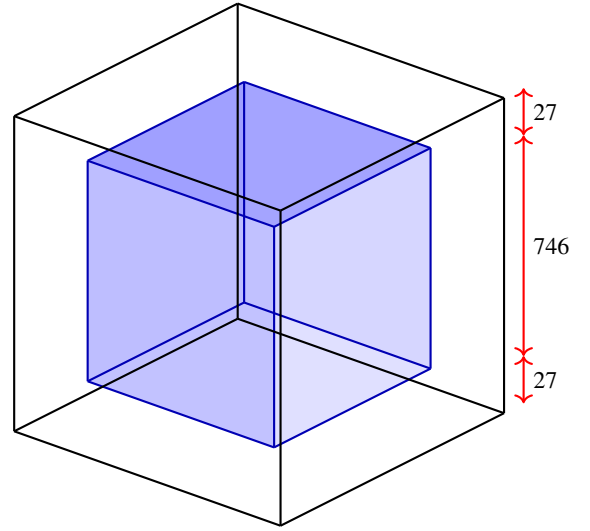

The governing equations are described in \cite{meng2020minimod}. The key element -and most computationally demanding- is the Laplacian found in the spatial differential operator. 
%The constant-density acoustic isotropic wave equation reads
%\begin{equation}
%  \frac{1}{V^2}\frac{\partial^2 u}{\partial t^2} - \nabla^2 u = f,
%  \label{eq:wave}
%\end{equation}
%where $u = u(x,y,z,t)$ is the presure wavefield, $V = V(x,y,z)$
%is the velocity of the pressure wavefield, and $f$ is the source perturbation.
%Discretising in time with a second-order centred finite-difference
%scheme yields the update equation
%\begin{equation}
%  \begin{split}
%    u^{n+1} & - Q\,u^{n} + u^{n-1} = \Delta t^{2}\,V^{2}\,f^{n}, \\
%    \textrm{with } Q &= 2 + \Delta t^{2}\,V^{2}\,\nabla^{2},
%  \end{split}
%  \label{eq:update}
%\end{equation}
%where $\Delta t$ is the time step and the superscript $n$ denotes
%the time level.
The Laplacian $\nabla^2$ is discretized with an
8th-order star-shaped stencil, giving a 25-point kernel in 3-D:
\begin{align}
  \nabla^2 u \approx \sum_{m=1}^{4} \Bigl[
    &c_x^{m}\bigl(u_{i+m,j,k}+u_{i-m,j,k}\bigr) \nonumber\\
    &+ c_y^{m}\bigl(u_{i,j+m,k}+u_{i,j-m,k}\bigr) \nonumber\\
    &+ c_z^{m}\bigl(u_{i,j,k+m}+u_{i,j,k-m}\bigr) \Bigr]
  \label{eq:stencil}
\end{align}
where $c_x^m$, $c_y^m$, $c_z^m$ are discretization parameters along each
axis respectively. The computational domain has two distinctive segments, the inner region (which represents the physical domain) and an auxiliary surrounding shell (PML boundary region) where a boundary condition is applied, as can be seen in Figure~\ref{fig:domain}.

\textbf{Inner region.}
In the interior of the domain, the solver evaluates the wave equation  using the 25-point
stencil~\eqref{eq:stencil} applied to the pressure field stored
in the \texttt{u}-array.
The seismic source is injected as a Gaussian wavelet at a fixed
grid point, providing a physically realistic, band-limited
excitation of the wavefield.
The multi-statement nature of this kernel, involving auxiliary
arrays for damping and source terms.

\textbf{PML boundary region.}
To suppress spurious reflections at the domain boundaries, we
apply a Perfectly Matched Layer (PML) condition~\cite{komatitsch2007unsplit}.
The PML region surrounds the inner cubic domain and is subdivided
into six sub-regions (top, bottom, front, back, left, right),
each processed by a dedicated kernel to avoid branch divergence.
The PML width of 27 cells is set by the physical parameters of
the seismic model: specifically, the absorbing layer must span
approximately three wavelengths at the dominant source frequency
(25\,Hz) to suppress reflections, yielding $3 \lambda_{\mathrm{max}} / \Delta x
= 3 \times (4500\,[\mathrm{m/s}] / 25\,[\mathrm{Hz}]) / 20\,[\mathrm{m}] = 27$ cell, where $\Delta x$ is the lattice spacing of the spatial discretization (identical in every direction)

This surrounding shell accounts for roughly 19\% of the total
$800^3$ grid\footnote{The memory allocated for the grid is $808^3$ to contain the halo.} volume (maximum that fit in the target GPU), leaving an inner region of $746^3$ cells for the primary wavefield computation.
In the PML region, again a 25-point stencil is applied.
The PML region and its kernel are not the bottleneck of the simulation, as they become negligible for larger grids (production).

\textbf{Domain decomposition.}
Following~\cite{sai2020}, we launch separate GPU kernels for the
inner region and each of the six PML sub-regions.
This separation eliminates boundary branch divergence and ensures
work balance within each kernel.
In contrast to~\cite{sai2020}, the kernels are launched
sequentially rather than concurrently: we tested concurrent PML
dispatch but found no throughput improvement on the B580,
as the inner kernel alone fully saturates the available execution
resources. Figure~\ref{fig:domain} illustrates the decomposition.

% \textbf{Double-buffer time stepping.}
% The time-stepping loop~\eqref{eq:update} requires the pressure field
% at two consecutive time levels, $u^{n}$ and $u^{n-1}$, to produce
% $u^{n+1}$.
% Rather than maintaining three separate arrays, we allocate two
% buffers of size $800^3$ and swap their roles after each step:
% the buffer that held $u^{n-1}$ is overwritten with $u^{n+1}$, and
% the pointers are exchanged so that the next iteration sees the
% correct $u^{n}$ and $u^{n-1}$.
% This ping-pong scheme eliminates an extra copy and simplifies the
% loop body, but it doubles the memory footprint of the pressure
% field compared to an in-place update, a trade-off that must be
% accounted for when sizing the problem to fit device
% memory~\cite{micikevicius2009,sai2020}.

\textbf{Buffer initialization and memory compression.}
The main floating-point buffers, must be allocated and initialized before the time-stepping
loop begins.
We evaluate two initialization strategies:
\begin{itemize}
  \item \emph{Zero initialization}: all buffers are filled with zero except for the source of Gaussian noise 
        that injects information at every steps in the inner domain (common choice in seismic simulation)
  \item \emph{Random initialization}: all buffers are filled with
        uniformly distributed random values.
\end{itemize}
This distinction is deliberately chosen to expose and quantify the
effect of the \emph{memory compression} feature present in the
Intel Battlemage architecture.
The BMG memory subsystem applies lossless compression directly from/to
main memory, reducing the effective number of bytes transferred
over the memory bus and thus increasing usable bandwidth,
a hardware compression at no software cost
\textit{e.g.} the measured bandwidth will be 100 GB/s but 
the ``dat'' transferred will be 800 GB/s.
Zero-initialized buffers are highly compressible (ratio 512) and therefore
benefit fully from this feature, delivering a significant throughput
boost that can be considered a \emph{free lunch}: no code change is
required, the hardware handles it transparently.

For the stencils problem with zero initialization is that 
as the simulation progresses 
and the information propagates through the
domain, the buffers gradually fill with non-trivial floating-point
data, reducing their compressibility until the gain eventually
vanishes.  Random-initialized buffers are incompressible from the start and
represent the worst-case scenario.

\begin{figure}
 \centering
 \includegraphics[height=3in,width=3.4in]{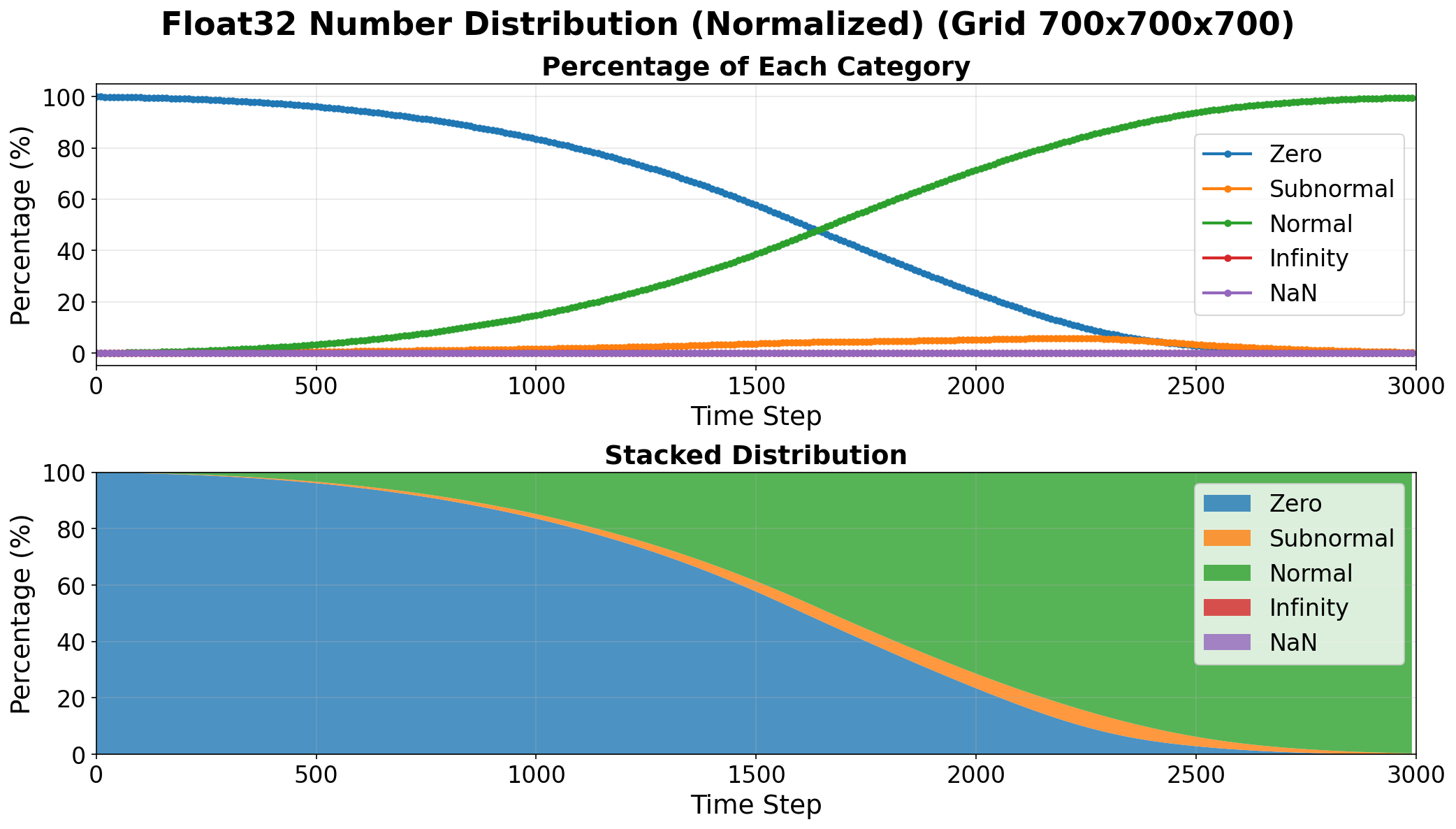}
 \caption{Distribution of floating-point values in the $700^3$ grid over time steps for a
 zero constant initialization.}
 \label{fig:numbers}
\end{figure}

This behavior is clearly visible in Figure~\ref{fig:numbers}, which shows the distribution of all numbers
(normalized/denormalized/zero/NaN) in the grid over the time iteration steps.

\subsection{TinyTC Tile Compiler}

TinyTC~\cite{tinytc} is a domain-specific language and compiler for
tensor computations targeting Intel GPU architectures that exposes the
hardware tiling capacity through a DSL (specific load/store capacity).
Its kernel language is written directly in \emph{Static Single
Assignment} (SSA) form, there is no high-level language front-end;
the programmer authors IR by hand (like CUDA Tile IR for NVIDIA).
%The IR draws heavily from MLIR and LLVM in both look and feel.
%Following LLVM conventions, global identifiers start with
%\texttt{@} and local identifiers with \texttt{\%}.
At compile time, TinyTC lowers the tensor IR
to SPIR-V, making it in effect a compact ($\approx$2\,MB)
LLVM-style front-end embeddable in any project that supports
OpenCL, Level Zero, or SYCL without pulling in heavy framework
dependencies.
It follows a \emph{Single Program Multiple Data} (SPMD) execution
model: the programmer explicitly partitions work across [a tile|tiles]. 
Compared to Triton, this gives finer control over the generated code
at the cost of greater verbosity and a steeper learning curve.

The central abstraction is the \emph{cooperative operation}:
the programmer specifies the shape, stride, and element type 
of a tile within a multidimensional tensor. 
TinyTC uses this information to emit \emph{2D block load} instructions 
that transfer a rectangular memory region in a single 
hardware operation to the LSC. These instructions are 
significantly more efficient than scalar or 1D-vectorized memory accesses.
Last and not least the 2D block load is \emph{hardware boundary
safety}: the instruction specification guarantees that accesses are
clamped to the declared buffer extents, making out-of-bounds reads
and writes architecturally impossible. As illustrated in the next listing.
Data is loaded with boundary check using
\texttt{cooperative\_matrix\_load.both\_checked}, which takes a 2D
\texttt{subview} and a row/column offset; out-of-bounds accesses are
clamped by hardware with no mask logic required.
Results are written back analogously:
\begin{lstlisting}
; 2D slice of 3D buffer at X-plane %lk
%u_xk = subview %u[0:%z_end, 0:%y_end, %lk]
; 2D block load, OOB clamped by hardware
%tile = cooperative_matrix_load.both_checked
            %u_xk[%z_off, %y_off] : $mat_t
...
; 2D block store
cooperative_matrix_store.both_checked
            %result, %v_xk[%z_off, %y_off]
\end{lstlisting}

The programmer therefore needs no mask logic and no padding: the
hardware automatically enforces the domain boundary at no cost.
On supported architectures, Triton generates 2D block loads 
%(although this is not currently supported on Battlemage, where it emits 
%sequential loads). 
However, these optimizations are applied transparently,
without giving the programmer direct control over the operation shape or the 
prefetch schedule. In contrast, TinyTC exposes these mechanisms, allowing the 
memory access pattern to be tuned to match the stencil’s reuse footprint.

%To conclude, the connection with SYCL is established as follows:
In the SYCL case,
the compiler takes the kernel IR as a plain C string,
compiles it to SPIR-V, and packages the result into a standard SYCL kernel bundle.
The resulting \texttt{sycl::kernel} object can then be dispatched using a standard 
\texttt{parallel\_for}, allowing TinyTC kernels to seamlessly integrate
into existing SYCL host code.
\begin{lstlisting}[language=C++]
// compile IR string -> SPIR-V -> SYCL kernel bundle
auto prog   = tinytc::parse_string(ir_source);
auto bundle = tinytc::create_kernel_bundle(
    queue.get_context(), queue.get_device(), prog,
    tinytc_core_feature_flag_large_register_file);
auto kernel = tinytc::create_kernel(bundle, "sc_inner_point");
// dispatch as a standard SYCL nd_range kernel
q.submit([&](sycl::handler &h) {
    h.set_args(x_begin, ..., u, u_shape0, ..., coef0);
    h.parallel_for(
        sycl::nd_range{global_range, group_size}, kernel);
});
\end{lstlisting}

\subsection{Triton}

Triton~\cite{triton} is a tile-based Python DSL in which the
programmer reasons about \emph{blocks} of data rather than individual
elements.
A kernel is launched over a grid of program instances, each
responsible for one tile of the output.
Within a kernel, data is brought into registers using
\texttt{tl.load}, which takes a block of pointers and an optional
boolean \emph{mask}; elements for which the mask is \texttt{False}
are replaced by a user-supplied fill value instead of triggering an
out-of-bounds access.
Symmetrically, \texttt{tl.store} writes a block back to memory,
again guarded by a mask.

This mask mechanism is the idiomatic way to handle domain boundaries:
the programmer constructs index tensors for the full tile, computes a
validity mask from the domain extents, and passes it to every load
and store.
No explicit branching or loop peeling is required; the compiler
lowers the masked operations to predicated instructions on the target
architecture.

Iteration over a dimension is expressed as a Python \texttt{for} loop
over a \texttt{tl.range}, with the loop body operating on the current
block of pointers.
The compiler is responsible for scheduling, register allocation, and
(on supported backends) software pipelining of the resulting memory
operations. 

To conclude, triton generates 2D block loads when tensor descriptor is included in the main code. However,
examination of the assembly does not reveal their use, resulting in lower performance (see section \ref{sec:results} for explanation).

% This model maps naturally onto stencil computations: the spatial
% axes of the grid become the tile dimensions, successive planes along
% the streaming axis are loaded in a loopr., and the halo points needed
% by the stencil are included in an enlarged tile or loaded
% separately before the main loop body.

% ---------------------------------------------------------------
\section{Implementation}
\label{sec:impl}

Both kernels implement the same 2.5D register-tiling
strategy~\cite{micikevicius2009,sai2020}.
The domain is partitioned into $TX \times TY \times TZ$ work-group
tiles; each work-group is responsible for one tile and sweeps it along
the X-axis.

\textbf{Tile decomposition.}
The 3D grid is indexed in Z-fastest (column-major) order.
Given a work-group identified by $(g_x, g_y, g_z)$, the tile origins
are $x_0 = g_x \cdot TX$, $y_0 = g_y \cdot TY$, $z_0 = g_z \cdot TZ$.
The Y/Z extent of the tile ($TY \times TZ$ points) fits in the LSC 
and is treated as a 2D matrix throughout (typically $TX=32$, $TY=16$, $TZ=16$
for a total of 32 KB).

\textbf{X register queue.}
Before the X-loop, the nine $TY \times TZ$ planes at
$x_0-4, \ldots, x_0+4$ are loaded into the LSC.
At each step $k$ the four-order X Laplacian is assembled
entirely from these LSC/register values, incurring no memory traffic.
After the update the oldest plane is discarded and a new plane at
$k{+}5$ is loaded, keeping the queue current for the next iteration.

\textbf{Y and Z stencils.}
For each step $k$ the eight $TY\times TZ$ planes at offsets
$\pm 1\ldots\pm 4$ along Y and Z are fetched from global memory.
No explicit cache management is required: the LSC
provides sufficient reuse across adjacent
tiles so that most of these loads are cache hits in practice.

More specifically the TinyTC implementation uses 2D block loads 
to fetch the Y/Z planes with halo. For the Triton version, the Y/Z 
planes are loaded without the halo.

The total memory traffic per output point is dominated by the forward
X-plane load (amortized over $TX$ steps). For TinyTC, each Y/Z plane 
is extended by 8 cells (±4) in both Y and Z dimensions to include the 
halo. For Triton, the Y/Z planes are loaded at the tile size without 
halo extension, and boundary halos are handled separately.

\begin{figure}[t]
  \centering
  \tdplotsetmaincoords{70}{110}
  \begin{tikzpicture}[tdplot_main_coords, scale=0.92, >=Stealth]
    % Domain extents (illustrative scale)
    \def\DX{7.0}
    \def\DY{4.0}
    \def\DZ{4.0}
    % Current k plane (mid-domain)
    \def\kk{3.5}
    % Tile size and origin on the YZ plane
    \def\tileY{2.0}
    \def\tileZ{2.0}
    \def\yorg{1.0}
    \def\zorg{1.0}
    % Tile centre (for arrows)
    \pgfmathsetmacro\yctr{\yorg+\tileY/2}
    \pgfmathsetmacro\zctr{\zorg+\tileZ/2}
    \pgfmathsetmacro\ymax{\yorg+\tileY}
    \pgfmathsetmacro\zmax{\zorg+\tileZ}

    % --- back/bottom fills drawn first ---
    \fill[gray!10,opacity=0.5]
      (0,0,0)--(\DX,0,0)--(\DX,\DY,0)--(0,\DY,0)--cycle;
    \fill[gray!10,opacity=0.5]
      (0,0,0)--(\DX,0,0)--(\DX,0,\DZ)--(0,0,\DZ)--cycle;

    % --- register-queue planes (4 ghost planes) ---
    \foreach \xi in {1.0, 2.2, 4.8, 6.0} {
      \fill[blue!18,opacity=0.55]
        (\xi,0,0)--(\xi,\DY,0)--(\xi,\DY,\DZ)--(\xi,0,\DZ)--cycle;
      \draw[blue!35,very thin]
        (\xi,0,0)--(\xi,\DY,0)--(\xi,\DY,\DZ)--(\xi,0,\DZ)--cycle;
    }

    % --- current active plane (k) ---
    \fill[blue!35,opacity=0.65]
      (\kk,0,0)--(\kk,\DY,0)--(\kk,\DY,\DZ)--(\kk,0,\DZ)--cycle;
    \draw[blue!65,thin]
      (\kk,0,0)--(\kk,\DY,0)--(\kk,\DY,\DZ)--(\kk,0,\DZ)--cycle;

    % --- active TZ x TY tile (orange) ---
    \fill[orange!60,opacity=0.88]
      (\kk,\yorg,\zorg)--(\kk,\ymax,\zorg)--
      (\kk,\ymax,\zmax)--(\kk,\yorg,\zmax)--cycle;
    \draw[orange!85,thick]
      (\kk,\yorg,\zorg)--(\kk,\ymax,\zorg)--
      (\kk,\ymax,\zmax)--(\kk,\yorg,\zmax)--cycle;

    % --- Z-neighbour load arrows (red) ---
    \pgfmathsetmacro\zarrtop{\zorg+\tileZ+1.1}
    \pgfmathsetmacro\zarrbot{\zorg-1.1}
    \draw[->,red!65,thick]
      (\kk,\yctr,\zctr)--(\kk,\yctr,\zarrtop);
    \draw[->,red!65,thick]
      (\kk,\yctr,\zctr)--(\kk,\yctr,\zarrbot);

    % --- Y-neighbour load arrows (dark green) ---
    \pgfmathsetmacro\yarrtop{\yorg+\tileY+1.1}
    \pgfmathsetmacro\yarrbot{\yorg-1.1}
    \draw[->,green!55!black,thick]
      (\kk,\yctr,\zctr)--(\kk,\yarrtop,\zctr);
    \draw[->,green!55!black,thick]
      (\kk,\yctr,\zctr)--(\kk,\yarrbot,\zctr);

    % --- X sweep arrow ---
    \draw[->,purple!75,very thick]
      (0.3,\yctr,\zctr)--(\DX-0.3,\yctr,\zctr);

    % --- domain wireframe (front edges) ---
    \draw[gray!55,thin]
      (0,0,\DZ)--(\DX,0,\DZ)--(\DX,\DY,\DZ)--(0,\DY,\DZ)--cycle;
    \draw[gray!55,thin]
      (0,\DY,0)--(\DX,\DY,0)--(\DX,\DY,\DZ)--(0,\DY,\DZ)--cycle;
    \draw[gray!55,thin]
      (0,0,0)--(0,\DY,0)  (0,0,0)--(0,0,\DZ)  (\DX,0,0)--(\DX,0,\DZ)
      (\DX,\DY,0)--(\DX,\DY,\DZ)  (\DX,0,\DZ)--(\DX,\DY,\DZ);

    % --- axis labels ---
    \draw[->,gray!70] (\DX+0.15,0,0)--(\DX+0.75,0,0)
      node[right,font=\scriptsize]{$X$};
    \draw[->,gray!70] (0,\DY+0.15,0)--(0,\DY+0.7,0)
      node[right,font=\scriptsize]{$Y$};
    \draw[->,gray!70] (0,0,\DZ+0.15)--(0,0,\DZ+0.7)
      node[above,font=\scriptsize]{$Z$};

    % --- text labels ---
    \node[blue!65,font=\scriptsize,anchor=south east]
      at (1.6,\DY,\DZ*0.7) {reg.~queue};
    \node[orange!75,font=\scriptsize,anchor=west]
      at (\kk+0.1,\ymax+0.1,\zmax+0.15) {$TZ\!\times\!TY$};
    \node[red!60,font=\scriptsize,anchor=west]
      at (\kk+0.1,\yctr+0.05,\zarrtop) {Z};
    \node[green!55!black,font=\scriptsize,anchor=south]
      at (\kk+0.1,\yarrtop,\zctr) {Y};
    \node[purple!70,font=\scriptsize,above]
      at (\DX*0.5,\yctr+0.25,\zctr) {X sweep};
  \end{tikzpicture}
  \caption{Register-queue tile decomposition. Blue planes are the nine
    $TZ\times TY$ tiles kept in registers (four shown). The orange panel
    is the active output tile at step~$k$. Red and green arrows mark
    the Z- and Y-direction loads fetched from global memory each step;
    the purple arrow shows the X sweep direction.}
  \label{fig:tile}
\end{figure}
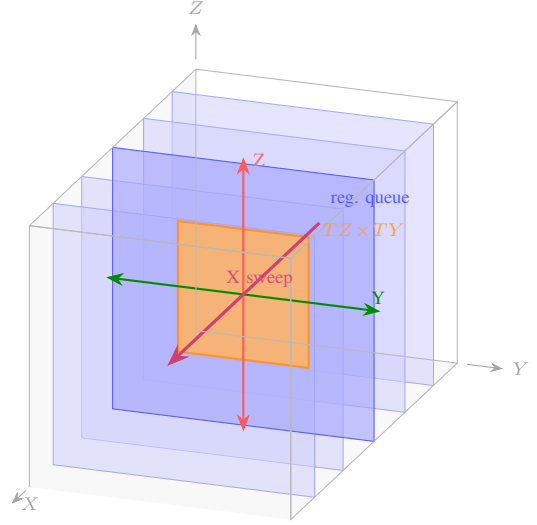

Listing~\ref{lst:algo} shows the loop structure as Triton Python pseudocode.
Figure~\ref{fig:tile} illustrates the decomposition in 3D.

\begin{lstlisting}[language=Python,
  caption={Register-queue algorithm (pseudocode).},
  label={lst:algo}]
# workgroup tile: z0:z0+TZ, y0:y0+TY, x0:x0+TX
z = z0 + arange(TZ)  # TZ-element vector
y = y0 + arange(TY)  # TY-element vector

# pre-fill X register queue: planes x0-4 .. x0+4
q = [u[z, y, x0+d] for d in range(-4, 5)]  # 9 tiles

for k in range(x0, x0 + TX):
    # X Laplacian: pure register arithmetic, no loads
    lapx = sum(cx[m]*(q[4-m]+q[4+m]) for m in 1..4)
    # Y/Z Laplacians: 8+8 global loads per step
    lapy = sum(cy[m]*(u[z,y-m,k]+u[z,y+m,k]) for m in 1..4)
    lapz = sum(cz[m]*(u[z-m,y,k]+u[z+m,y,k]) for m in 1..4)
    # leapfrog update
    v[z,y,k] = 2*u[z,y,k] - v[z,y,k] + \
        roc2[z,y,k]*(c0*u[z,y,k]+lapx+lapy+lapz)
    # advance queue: pop front, push next plane
    q = q[1:] + [u[z, y, k+5]]
\end{lstlisting}

\subsection{TinyTC Kernel}

Listing~\ref{lst:tinytc} shows a condensed excerpt from the inner
stencil kernel written in the TinyTC IR language.
The \texttt{shape\_gcd}/\texttt{stride\_gcd} annotations on each
\texttt{memref} argument let the compiler prove that the base pointer
and strides satisfy the 4-, 16-, and 64-byte alignment requirements
of the 2D block load instruction.
\texttt{foreach\_tile} distributes the $TZ\times TY$ work-group
tile into $\$tz\times\$ty$ sub-tiles, one per subgroup.
Inside the X-loop, \texttt{subview} selects a 2D slice, the
Z- and Y-direction neighbors are loaded with
\texttt{cooperative\_matrix\_load.both\_checked}, the Laplacian is
accumulated via \texttt{cooperative\_matrix\_scale} and element-wise
\texttt{add}, and the result is written back with
\texttt{cooperative\_matrix\_store.both\_checked}.
The X-direction is handled without extra loads via a register queue
pre-filled before the loop and shifted by one plane per iteration.

\begin{lstlisting}[caption={Condensed TinyTC IR for the inner stencil
kernel (Z-direction contribution shown).},label={lst:tinytc}]
func @sc_inner_point(
  %u: memref<f32x?x?x?,strided<1,?,?>>
      {shape_gcd=[4,4,4],stride_gcd=[1,4,16]},
  %v: memref<f32x?x?x?,strided<1,?,?>>
      {shape_gcd=[4,4,4],stride_gcd=[1,4,16]},
  %roc2: memref<f32x?x?x?,strided<1,?,?>>,
  ..., %coef0: f32)
  attributes {subgroup_size=16,
              work_group_size=[16,2]} {
  ...
  foreach_tile (%i,%j)=(%c0,%c0),(%TZ,%TY)
              as (%ti,%tj) <= ($tz,$ty) {
    ; pre-fill X-queue: planes x-4 .. x+4 in registers
    ...
    for %k=%x_begin,%x_end init(...) -> (...) {
      ; 2D slice at X-plane %lk
      %u_xk = subview %u[%c0:%lz_end,%c0:%ly_end,%lk]
                      : memref<f32x?x?,strided<1,?>>
      ; X-stencil computed from register queue (no loads)
      %lapx = ...
      ; Z stencil: 2D block loads, OOB clamped by hardware
      %z_1 = cooperative_matrix_load.both_checked
                %u_xk[%ltile_z_1,%ltile_y] : $mat_t
      %z1  = cooperative_matrix_load.both_checked
                %u_xk[%ltile_z1, %ltile_y] : $mat_t
      %lapz10 = add %z1, %z_1 : $mat_t
      %lapz11 = cooperative_matrix_scale %cr1z, %lapz10 : $mat_t
      %lapz = add %lapz11, ... : $mat_t
      ; Y stencil: same pattern as Z
      %lapy = ...
      %stencil = add %lapx, (add %lapy, %lapz) : $mat_t
      ; assemble update: 2*u - v + roc2*stencil
      %result = add (sub (cooperative_matrix_scale 2.0, %x0), %vv),
                    (mul %r2, %stencil) : $mat_t
      ; write result tile to v
      cooperative_matrix_store.both_checked
                %result, %v_xk[%ltile_z,%ltile_y]
      ; advance register queue
      yield (%x_3,%x_2,%x_1,%x0,%x1,%x2,%x3,%x4,%x5)
    }
  }
}
\end{lstlisting}

\subsection{PT/Triton Kernel}

The Triton inner kernel follows the same register-queue algorithm
described above (see Listing~\ref{lst:tinytc}).
However, Triton's masked \texttt{tl.load} carries a measurable
overhead in 2D tiling scenarios: generating a per-element predication
mask across the full $BLOCK\_Y \times BLOCK\_Z$ tile and threading it
through every neighbor load adds instruction pressure that degrades
performance, especially when only the last tile along each axis is
partially out-of-bounds.

The adopted solution is to \emph{shift the boundary check entirely to
the store side} and to ensure loads are always safe by construction
(Fig.~\ref{fig:padding}).
All buffers (\texttt{u}, \texttt{v}, \texttt{phi}, \texttt{eta})
are allocated with one extra tile of padding beyond the physical
domain on each axis:
\begin{lstlisting}[language=Python]
pad_y, pad_z = block_y, block_z
u = torch.zeros((nx + 2*lx + pad_x,
                 ny + 2*ly + pad_y,
                 nz + 2*lz + pad_z), ...)
\end{lstlisting}
The kernel grid is launched with \texttt{triton.cdiv}, so the last
tile along each direction may extend one tile beyond the valid domain.
Because the padding is at least one full tile wide, all loads
including the stencil halo at $\pm 4$ and the pre-fill of the
X-queue at $x_0 \pm 4$ land in allocated, zeroed memory and
require no mask:
\begin{lstlisting}[language=Python]
# unmasked load: always in-bounds due to padding
y_p1 = tl.load(u_ptr + (x+lx)*sx + (y_l+1)*sy + z_l*sz)
\end{lstlisting}
The store, on the other hand, must not write computed values into the
padding region; incorrect values there would corrupt subsequent
time-step loads.
A validity mask is therefore applied exclusively at the store:
\begin{lstlisting}[language=Python]
mask_out = (y < y_end) & (z < z_end) & (x < x_end)
tl.store(v_ptr + ..., out, mask=mask_out)
\end{lstlisting}

\begin{figure}[t]
  \centering
  \begin{tikzpicture}[>=Stealth, font=\scriptsize]
    % cell width
    \def\s{0.5}

    % --------------------------------------------------------
    % (a) WITHOUT PADDING  (domain=10, block=4)
    % Cells 10,11 are out-of-bounds
    % --------------------------------------------------------
    \foreach \i in {0,...,9} {
      \draw[fill=blue!10, draw=gray!40]
        ({\i*\s}, 0) rectangle ({(\i+1)*\s}, 0.38);
    }
    \foreach \i in {10,11} {
      \draw[fill=red!20, draw=red!40]
        ({\i*\s}, 0) rectangle ({(\i+1)*\s}, 0.38);
      \node[red!65, font=\tiny] at ({(\i+0.5)*\s}, 0.19) {$\times$};
    }
    % domain boundary
    \draw[black!70, thick, dashed] ({10*\s},-0.12)--({10*\s},0.52);
    % tile outlines
    \draw[blue!70, thick, rounded corners=1.5pt]
      (0,-0.07) rectangle ({4*\s},0.45);
    \draw[teal!70!black, thick, rounded corners=1.5pt]
      ({4*\s},-0.07) rectangle ({8*\s},0.45);
    \draw[orange!80!black, thick, rounded corners=1.5pt]
      ({8*\s},-0.07) rectangle ({12*\s},0.45);
    % tile labels
    \node[blue!70, anchor=south]         at ({2*\s},0.45) {\tiny tile\,0};
    \node[teal!70!black, anchor=south]   at ({6*\s},0.45) {\tiny tile\,1};
    \node[orange!80!black, anchor=south] at ({10*\s},0.45) {\tiny tile\,2};
    % brace below tile 2
    \draw[red!55,thin] ({8*\s},-0.18)--({8*\s},-0.30)--({10*\s},-0.30);
    \draw[red!55,thin] ({12*\s},-0.18)--({12*\s},-0.30)--({10*\s},-0.30);
    \node[red!65, anchor=north] at ({10*\s},-0.30)
      {\tiny mask on every load};
    % panel label + axis
    \node[anchor=east] at (-0.12,0.19) {(a)};
    \node[anchor=west] at ({12*\s+0.1},0.19) {\tiny $Z$};

    % --------------------------------------------------------
    % (b) WITH ONE-TILE PADDING  (pad=4 extra cells: 10..13)
    % --------------------------------------------------------
    % row offset
    \def\yb{-1.45}
    \pgfmathsetmacro\ybhi{\yb+0.38}
    \foreach \i in {0,...,9} {
      \draw[fill=blue!10, draw=gray!40]
        ({\i*\s},\yb) rectangle ({(\i+1)*\s},{\yb+0.38});
    }
    \foreach \i in {10,...,13} {
      \draw[fill=gray!25, draw=gray!35]
        ({\i*\s},\yb) rectangle ({(\i+1)*\s},{\yb+0.38});
    }
    % domain boundary
    \pgfmathsetmacro\ybext{\yb-0.12}
    \pgfmathsetmacro\ybhi{\yb+0.50}
    \draw[black!70, thick, dashed] ({10*\s},\ybext)--({10*\s},\ybhi);
    % tile outlines
    \pgfmathsetmacro\yblo{\yb-0.07}
    \pgfmathsetmacro\ybup{\yb+0.45}
    \draw[blue!70, thick, rounded corners=1.5pt]
      (0,\yblo) rectangle ({4*\s},\ybup);
    \draw[teal!70!black, thick, rounded corners=1.5pt]
      ({4*\s},\yblo) rectangle ({8*\s},\ybup);
    \draw[orange!80!black, thick, rounded corners=1.5pt]
      ({8*\s},\yblo) rectangle ({12*\s},\ybup);
    % tile labels
    \node[blue!70, anchor=south]         at ({2*\s},{\ybup+0.09}) {\tiny tile\,0};
    \node[teal!70!black, anchor=south]   at ({6*\s},{\ybup+0.09}) {\tiny tile\,1};
    \node[orange!80!black, anchor=south] at ({10*\s},{\ybup+0.09}) {\tiny tile\,2};
    % store markers inside valid cells (8,9) and circles in padding (10,11)
    \pgfmathsetmacro\ybmid{\yb+0.19}
    \foreach \i in {8,9} {
      \node[green!55!black, font=\normalsize] at ({(\i+0.5)*\s},\ybmid)
        {\checkmark};
    }
    \foreach \i in {10,11} {
      \node[gray!55] at ({(\i+0.5)*\s},\ybmid) {$\circ$};
    }
    % padding span arrow ABOVE the padding cells
    \pgfmathsetmacro\ybpad{\yb+0.55}
    \draw[<->, gray!55, thin]
      ({10*\s},\ybpad)--({14*\s},\ybpad)
      node[midway, above, gray!65] {\tiny 1 tile pad};
    % annotation below
    \pgfmathsetmacro\ybann{\yb-0.18}
    \node[green!55!black, anchor=north] at ({10*\s},\ybann)
      {\tiny loads free~$|$~store masked at $k{<}N$};
    % panel label + axis
    \node[anchor=east] at (-0.12,\ybmid) {(b)};
    \node[anchor=west] at ({14*\s+0.08},\ybmid) {\tiny $Z$};

    % legend
    \def\yl{-2.35}
    \draw[fill=blue!10, draw=gray!40] (0.00,\yl) rectangle (0.28,{\yl+0.22});
    \node[anchor=west] at (0.34,{\yl+0.11}) {\tiny valid domain};
    \draw[fill=gray!25, draw=gray!35] (1.60,\yl) rectangle (1.88,{\yl+0.22});
    \node[anchor=west] at (1.94,{\yl+0.11}) {\tiny padding (zeroed)};
    \draw[fill=red!20, draw=red!40]   (3.60,\yl) rectangle (3.88,{\yl+0.22});
    \node[anchor=west] at (3.94,{\yl+0.11}) {\tiny OOB};

  \end{tikzpicture}
  \caption{Boundary handling along one axis ($Z$; block size~4,
    domain~10 cells). (a)~Without padding tile~2 spills into
    unallocated memory (red), requiring a predication mask on every
    load. (b)~One extra tile of zeroed memory is appended; all loads
    land in allocated space with no mask. The store uses a mask to
    avoid writing into the padding region
    ($\checkmark$~=~written, $\circ$~=~skipped).}
  \label{fig:padding}
\end{figure}
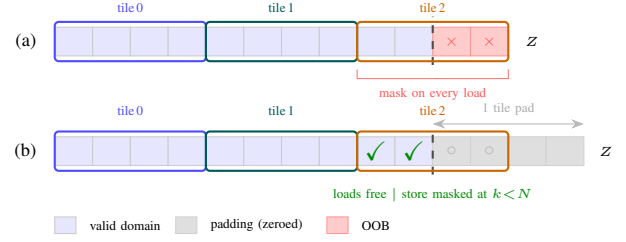

This ``load wide, store narrow'' strategy eliminates all per-load
predication at the cost of a small over-allocation (~one extra
$BLOCK\_Y \times BLOCK\_Z$ slice), which is negligible for the
800\textsuperscript{3} grids used in this evaluation.
The approach depicted in Fig.~\ref{fig:padding}(b) is the one
actually used in the PT/Triton implementation.

\subsection{SYCL Shared-Memory Kernel}
\label{sec:impl:sycl}

As a portable reference baseline, we implemented the stencil kernel
in standard SYCL following the shared-memory tiling strategy of
Micickevicius~\cite{micikevicius2009} for the inner kernel only.
The damping kernel, by contrast, uses a straightforward streaming
approach: data is read directly from main memory with no tiling or
data reuse optimization, reflecting the lower arithmetic intensity
of the PML update.
For the inner kernel, a $(BLOCK\_Y + halo) \times (BLOCK\_Z + halo)$ tile of the input grid is loaded
into \texttt{local\_accessor} shared memory, the work-group is synchronized,
by a per-thread X-direction register window.
No inline ASM for 2D block-load instructions are used: all global
memory accesses are scalar loads.

\subsection{OpenMP Offload Kernel}
\label{sec:impl:omp}

A second reference implementation uses OpenMP target offload,
offloading the standard 25-point kernel composed of three nested loops
to the GPU with minimal code changes.
The innermost loops are mapped onto GPU threads via
\texttt{\#pragma omp target teams distribute parallel for}, with no
explicit shared-memory or register-queue optimization.
This baseline represents the lowest-effort GPU port and provides a
lower bound on achievable throughput, useful for assessing how much
performance the more specialized tile-compiler kernels recover.

%\subsection{Claude and Triton}
%\label{sec:genai}
% \textcolor{red}{
%The rise of Generative AI opens new possibilities. Python, as the dominant 
%language in the AI ecosystem, combined with Triton, offers new perspectives 
%for kernel optimization. By defining strict validation tests, the stencil kernel described above
%is optimized iteratively with the help of Claude. 
%Already NVIDIA pushes this approach way forward with autonomous agent \cite{kernelagent2026}},
%but in this work, we prefer keep %control. 

%\todo[inline]{MA reviews please}

\section{Experimental Setup}
\label{sec:setup}

\textbf{Platforms.}
The primary evaluation target is the Intel Arc Battlemage~580
(B580), a consumer-grade discrete GPU with a theoretical peak memory
bandwidth of 450\,GB/s.
Although positioned as a gaming accelerator at a modest price point,
the B580 is the only publicly available host for the TinyTC tile
compiler at the time of writing and therefore represents the hardware
of primary interest.
The evaluation is extended to  higher-end platforms for
comparison: the Intel B70, GPU MAX 1550, NVIDIA A100/RTX6000
Blackwell/H100, and AMD MI325x.

The tinyTC runs are limited to Intel GPUs, whereas the PT/Triton runs perform on all GPUs. Table~\ref{tab:hw} summarizes the platforms and their main 
characteristics.

\begin{table}
  \centering
  \caption{Hardware platforms used in the evaluation. The Intel GPU Max~1550 number is given in 1-tile mode, i.e., with 64 Xe cores active instead of 128.}
  \label{tab:hw}
  \resizebox{\columnwidth}{!}{%
  \begin{tabular}{llccc}
    \multirow{2}{*}{\textbf{Hardware}}     & \multirow{2}{*}{\textbf{Architecture}} & \shortstack{\textbf{Peak BW Th.}} & \shortstack{\textbf{STREAM}} & \shortstack{\textbf{Peak Vec. FP32}}\\
    & & GB/s & GB/s & TFLOPs \\
    \midrule
    Intel B580         & X\textsuperscript{e}\,core2      & 450  & 400  & 13.7 \\
    Intel B70         & X\textsuperscript{e}\,core2       & 600  & 540  & 21.9 \\
    Intel GPU Max~1550 (1T)& X\textsuperscript{e}\,core1      & 1600 & 1000 & 21.2 \\
    NVIDIA A100      & Ampere    & 1935 & 1806 & 19.5  \\
    NVIDIA RTX6000   & Blackwell & 1597 & 1484 & 38.7  \\
    NVIDIA H100      & Hopper    & 2040 & 1917 & 51.2  \\
    AMD Mi325x       & CDNA 4    & 6000 & 4526 & 163.4 \\ 
  \end{tabular}}
\end{table}

%\todo[inline]{MA: need to add A100, RTX6K and Mi325x, including stream GPU}

\textbf{Memory compression on B580.}
The Battlemage memory subsystem implements lossless hardware
compression: cache lines containing repetitive or constant data
(e.g., all-zero patterns) are compressed on the fly by the memory
controller during transfer \textit{i.e.} GPU-DRAM to LSC, reducing the number of bytes that
traverse the memory bus and yielding an effective bandwidth exceeding
the theoretical peak without altering the data layout in main memory.
To characterize this effect quantitatively, experiments were conducted with a modified GPU
STREAM~\cite{stream} benchmark in which a controlled fraction of each array is
filled with uniformly distributed random values while the remainder
retains the STREAM defaults ($A{=}1$, $B{=}2$, $C{=}0$).
At a fill fraction of $1/128$ (nearly all constant data), the Triad
kernel reaches $\approx840$\,GB/s, nearly twice the theoretical
peak, while at full random fill ($1/1$) the effective bandwidth
converges to $\approx400$\,GB/s, consistent with the theoretical
limit.
Table~\ref{tab:stream_gpu_bw} reports the full sweep. This feature was enabled on the B580 and B70, not 
on the Intel GPU MAX~1550, and on the NVIDIA and AMD GPUs, its behavior was not characterized.

 Note that the behavior of Intel GPU in-memory compression can be controlled through the combination of two environment variables: \texttt{NEOReadDebugKeys=1} and \texttt{RenderCompressedBuffersEnabled=0|1}, however for this study, the memory compression was on for B580/B70 and off for Intel GPU MAX~1550.

\begin{table}[ht!]
  \centering
  \caption{STREAM GPU benchmark: effective memory bandwidth (GB/s)
    on the Intel B580 for varying data fill fractions (float32,
    $3\times 10$\,GB arrays).
    Fill fraction gives the proportion of elements initialized to
    random values in $[0,1]$; the remainder uses STREAM defaults
    ($A{=}1$, $B{=}2$, $C{=}0$). Results are the mean of 3 runs.}
  \label{tab:stream_gpu_bw}
  \setlength{\tabcolsep}{5pt}
  \begin{tabular}{l rrrr}
%    & \multicolumn{4}{c}{\textbf{float32 (GB/s)}} \\
    \textbf{Fill} & \textbf{Copy} & \textbf{Scale} & \textbf{Add} & \textbf{Triad} \\
    \midrule
    1/128 & 610.8 & 741.0 & 892.1 & 839.7 \\
    1/64  & 745.5 & 771.7 & 884.3 & 834.0 \\
    1/32  & 734.5 & 759.4 & 868.4 & 819.4 \\
    1/16  & 715.8 & 736.1 & 837.2 & 792.6 \\
    1/8   & 675.1 & 693.8 & 781.3 & 742.9 \\
    1/4   & 617.5 & 621.8 & 690.2 & 661.3 \\
    1/2   & 522.7 & 515.5 & 560.0 & 542.0 \\
    1/1   & 399.3 & 383.3 & 405.7 & 397.7 \\
  \end{tabular}
\end{table}

% ---------------------------------------------------------------
\section{Results}
\label{sec:results}

\subsection{Benchmark Results}

Table~\ref{tab:throughput_benchmark} presents the throughput (Giga points per second - Gpts/s) achieved by all
four implementations (Triton, TinyTC, SYCL, and OMP) across
1\,000, 2\,000, and 4\,000 time steps, with and without random field
initialization, for a computing domain fixed in $800^3$ grid points.

\begin{table}[t]
\centering
\caption{Throughput benchmark results (Gpts/s) for all implementations on Intel Arc B580 with 800$^3$ grid.}
\label{tab:throughput_benchmark}
\small
%\begin{tabular}{l|cc|cc|cc}
%\toprule
%\multirow{2}{*}{\textbf{Implementation}} & \multicolumn{2}{c|}{\textbf{1K Steps}} & \multicolumn{2}{c|}{\textbf{2K Steps}} & \multicolumn{2}{c}{\textbf{4K Steps}} \\
% & Random & ZeroInit & Random & ZeroInit & Random & ZeroInit \\
%\midrule
%TinyTC & 15.6 & 35.8 & 15.6 & 33.1 & 15.6 & 23.8 \\
%Triton & 13.5 & 32.5 & 13.6 & 30.3 & 13.6 & 21.5 \\
%SYCL   & 12.2 & 20.2 & 12.2 & 19.1 & 12.2 & 16.1 \\
%OMP    & 7.8  & 8.9  & 7.8  & 8.9  & 7.8  & 8.4 \\
%\bottomrule
%\end{tabular}
\begin{tabular}{l ccc ccc}
\multirow{2}{*}{\textbf{Version}} & \multicolumn{3}{c}{\textbf{Zero-Initialization}} & \multicolumn{3}{c}{\textbf{Random Initialization}} \\
 & 1k & 2k & 4k & 1k & 2k & 4k \\
\midrule
TinyTC  & 35.8 & 33.1& 23.8 & 15.6 & 15.6 & 15.6 \\
Triton  & 32.5 & 30.3 & 21.5 & 13.5 & 13.6 & 13.6 \\
SYCL    & 20.2 & 19.1 & 16.1 & 12.2 & 12.2 & 12.2 \\
OMP     & 8.9 & 8.9 & 8.4 & 7.8 & 7.8 & 7.8 \\
\end{tabular}
\end{table}
A first key observation is the dramatic performance difference between initialization strategies:
TinyTC throughput drops from 35.8\,Gpts/s (zero-initialization) to 15.6\,Gpts/s (random initialization),
a 56\% reduction; Triton drops from 32.5\,Gpts/s to 13.5\,Gpts/s (58\% reduction),
SYCL from 20.2\,Gpts/s to 12.2\,Gpts/s (39\% reduction), and OMP from 
8.9\,Gpts/s to 7.8\,Gpts/s (12\% reduction).
This discrepancy reflects the impact of Intel Battlemage's memory-compression
feature: zero-initialized arrays are highly compressible, yielding much higher
effective bandwidth and throughput, whereas random data is incompressible and
delivers throughput limited by the hardware's nominal bandwidth.

Another observation emerges from examining the step-count behavior.
With random initialization, all implementations show convergence to steady-state throughput
values: Triton stabilizes at 13.6~Gpts/s, TinyTC at 15.6~Gpts/s, SYCL at 12.2~Gpts/s,
and OMP at 7.8~Gpts/s (1000, 2000, and 4000 steps yield nearly identical throughput).
With zero-initialization, the throughput \emph{decreases} with more steps: TinyTC drops
from 35.8 to 23.8~Gpts/s, Triton from 32.5 to 21.5~Gpts/s, and SYCL from 20.2 to 16.1~Gpts/s.
This behavior indicates that random initialization prevents compression across all
time steps (constant nominal bandwidth), whereas zero-initialized arrays compress well initially but become
less compressible as structured wave patterns emerge and propagate through the domain. 
Ultimately, for a very large number of iterations, the throughput values converge between
the two initialization strategies. But due to the fast gain at the beginning, it takes time

%\begin{figure}[H]
%  \centering
%  \includegraphics[width=\columnwidth]{plot/noinit.png}
%  \caption{Throughput (Gpts/s) with zero-initialization
%           for 1\,000, 2\,000, and 4\,000 time steps.}
%  \label{fig:noinit}
%\end{figure}

%\begin{figure}[H]
%  \centering
%  \includegraphics[width=\columnwidth]{plot/init.png}
%  \caption{Throughput (Gpts/s) with random field initialization,
%           demonstrating improved performance due to memory compression.}
%  \label{fig:init}
%\end{figure}

%\subsection{Roofline Analysis}

\begin{figure}
  \includegraphics[height=3in,width=3.4in]{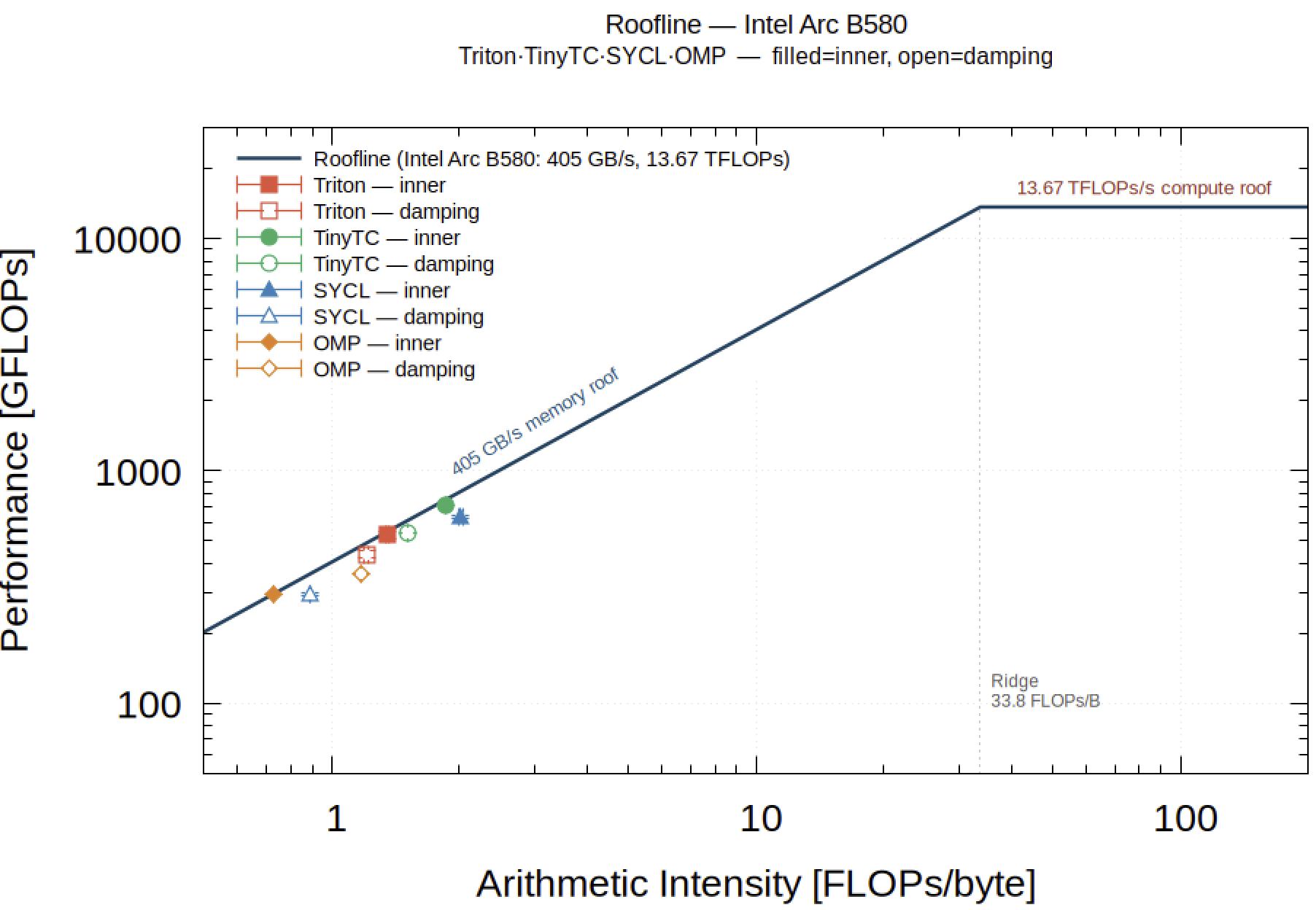}
  \caption{Roofline plot comparing all four implementations (TinyTC, Triton, SYCL, and OpenMP)
           on Intel Arc B580. The plot shows arithmetic intensity (FLOPs/byte) versus
           performance (GFLOPs/s) for both inner and damping kernels, with diagonal lines
           indicating memory bandwidth limits and the horizontal line showing the peak
           computational throughput.}
  \label{fig:roofline_b580}
\end{figure}

%\subsection{Discussion}

The benchmark results in Table~\ref{tab:throughput_benchmark} show that the
end-to-end throughput depends critically on the memory state: whether the
inputs arrays are initialized with constant data (best-case compression) or
contain random data (worst-case compression).

\subsection{Connecting Roofline to End-to-End Performance}

The roofline model was generated using unitrace, 
where the required counters were measured to determine bandwidth and FLOPs.

Figure~\ref{fig:roofline_b580} and Table~\ref{tab:roofline_b580} provide
insight into this difference by measuring the roofline performance of
individual inner and damping kernels under optimal (initialization) memory
compression conditions. 

All implementations are bandwidth-limited (lying below the memory ceiling in
Figure~\ref{fig:roofline_b580}), meaning that reducing the total memory traffic, and efficient caching is the primary path to higher throughput.

\begin{table}
  \centering
  \caption{Roofline metrics on the Intel Arc B580 (peak BW\,=\,405\,GB/s,
    peak Perf\,=\,13\,670\,GFLOPs/s).
    Values are the mean over 9 steady-state iterations; $\pm$ gives the standard deviation.
    AI\,=\,arithmetic intensity [FLOPs/byte].
    Time is the mean GPU execution time per kernel call (inner) or
    per 6$\times$ damping group (damping).}
  \label{tab:roofline_b580}
  \resizebox{\columnwidth}{!}{%
  \begin{tabular}{ll c cc cc cc}
    \textbf{Version} & \textbf{Kernel} &
      \textbf{AI} &
      \multicolumn{2}{c}{\textbf{Perf [GFLOPs/s]}} &
      \multicolumn{2}{c}{\textbf{BW [GB/s]}} &
      \multicolumn{2}{c}{\textbf{Time [ms]}} \\
    & & [FLOPs/B] & mean & $\pm$std & mean & $\pm$std & mean & $\pm$std \\
    \midrule
    \multirow{2}{*}{TinyTC}
      & inner   & 1.862 & 714.5 &  0.5 & 383.7 &  0.2 &  20.32 & 0.01 \\
      & damping & 1.511 & 539.7 &  0.9 & 357.1 &  0.5 &  13.83 & 0.02 \\
    \midrule
    \multirow{2}{*}{Triton}
      & inner   & 1.350 & 531.1 & 20.1 & 393.4 & 11.3 &  23.69 & 0.91 \\
      & damping & 1.207 & 436.1 & 15.4 & 361.4 &  9.0 &  15.09 & 0.54 \\
    \midrule
    \multirow{2}{*}{SYCL}
      & inner   & 2.015 & 634.2 &  9.7 & 314.8 &  5.0 &  26.13 & 0.40 \\
      & damping & 0.888 & 295.8 &  5.0 & 333.2 &  5.3 &  17.07 & 0.28 \\
   \midrule
    \multirow{2}{*}{OMP}
      & inner   & 0.729 & 293.3 &  0.2 & 402.5 &  0.2 &  45.99 & 0.03 \\
      & damping & 1.175 & 360.8 &  0.4 & 307.0 &  0.4 &  18.22 & 0.02 \\
  \end{tabular}}
\end{table}

The roofline analysis in Table~\ref{tab:roofline_b580} reveals that TinyTC's and Triton's
2D block-load strategy delivers bandwidth efficiency: sustaining $\approx$ 390~GB/s
for the inner kernel and $\approx 
 360$~GB/s for the damping kernel, close to the measurable STREAM bandwidth (405~GB/s).
TinyTC achieving the lowest DRAM traffic (5.9~GB \footnote{
In this section the roofline analysis of the kernel has been performed
on 10 iterations only. It explains the DRAM traffic.
}) and highest L3
hit rate (73.5\%) (see Table~\ref{tab:cache_b580}).
Triton achieves competitive bandwidth (393.4~GB/s for inner) and even better LSC hit rates
(70.0\%), but loses effectiveness due to scalar load emission.
However, the roofline measurements capture only the steady-state inner kernel
performance under optimal (initialization) conditions.

To understand the performance differences between implementations, we
instrumented the memory hierarchy with the unitrace \texttt{MemoryProfile}
counter set and extracted the LSC (L1 data cache), L3, and DRAM
traffic for each inner kernel under initialization conditions.
Table~\ref{tab:cache_b580} summarizes the results.

\begin{table}
  \centering
  \caption{Memory hierarchy metrics for the inner kernel on the Intel Arc B580
    (mean over 9 steady-state iterations).
    DRAM\,Rd: bytes read from main memory. LSC\,Rd: bytes read through the
    Load-Store Cache (L1). L3\,hit and LSC\,hit: hardware hit rates.}
  \label{tab:cache_b580}
  \resizebox{\columnwidth}{!}{%
  \begin{tabular}{l c c c c c}
    \textbf{Version} &
      \textbf{Time [ms]} &
      \textbf{DRAM Rd [GB]} &
      \textbf{LSC Rd [GB]} &
      \textbf{L3 hit [\%]} &
      \textbf{LSC hit [\%]} \\
    \midrule
    TinyTC & 20.3 &  \textbf{5.9} & 32.3 & \textbf{73.5} &           65.3 \\
    Triton & 23.7 &  7.4 & 27.7 &          61.3  & \textbf{70.0} \\
    SYCL   & 26.1 &  6.2 &  8.9 &          46.4  &          24.9 \\
    OMP    & 46.0 & 16.7 & 41.8 &          58.2  &          63.4 \\
  \end{tabular}}
\end{table}
The data confirm the hypothesis.
TinyTC achieves the lowest DRAM traffic (5.9\,GB) and the highest L3
hit rate (73.5\%), meaning that most of the stencil neighborhood
data is satisfied from the on-chip cache hierarchy rather than from
main memory.
This is a direct consequence of operating \emph{close to the metal}:
TinyTC exposes the Intel hardware tiling hierarchy through its cooperative
matrix API, and the programmer explicitly controls the shape,
alignment, and prefetch schedule of every 2D block load.
Inspection of the generated assembly confirms this: the TinyTC inner kernel
is dominated by \texttt{load2dstrided} (2D block-load) instructions,
which transfer a full rectangular tile from memory in a single
hardware operation, maximising spatial locality and allowing
a better usage of the L3 bandwidth.

Triton follows the same tile-based philosophy and achieves the best LSC hit rate
(70.0\%), but pays a higher DRAM cost (7.4 GB) and runs 14\% slower than TinyTC. Assembly inspection reveals the reason: our Triton implementation does 
\emph{not} use tensor descriptors, which are required to unlock 2D block-load 
instructions on Intel hardware. Without them, tile loads are lowered to scalar 
or 1D-vectorized loads, forgoing the bandwidth efficiency that 2D block loads 
provide. This was a deliberate simplification driven by time constraints rather 
than a fundamental compiler limitation: Triton \emph{can} emit 2D block loads 
when tensor descriptors are used explicitly. Exploiting tensor descriptors in 
the Triton kernel is left as future work and is expected to close the observed 
performance gap.

The SYCL shared-memory kernel routes data through the SLM scratchpad
rather than the LSC, which explains its very low LSC read volume (8.9\,GB)
and poor LSC hit rate (24.9\%).
While SLM staging avoids redundant global-memory fetches within a
work-group, it  produces a worse L3 hit rate (46.4\%) than either tile-compiler
approach.
In effect, the programmer-managed scratchpad competes with, rather
than complements, the hardware cache hierarchy on this architecture.
The SYCL results are not optimal, additional effort in manual tuning could have mitigated the issues. 

The current OMP offload kernel, with no data-reuse strategy, generates
nearly three times more DRAM traffic (16.7\,GB) than TinyTC and
takes more than twice as long (46\,ms vs.\ 20\,ms), providing the
expected lower bound on performance.

\subsection{Intel B70/GPUMAX 1550, NVIDIA A100/RTX6K/H100 and AMD mi325x}

\begin{figure}
 \centering
 \includegraphics[scale=0.23]{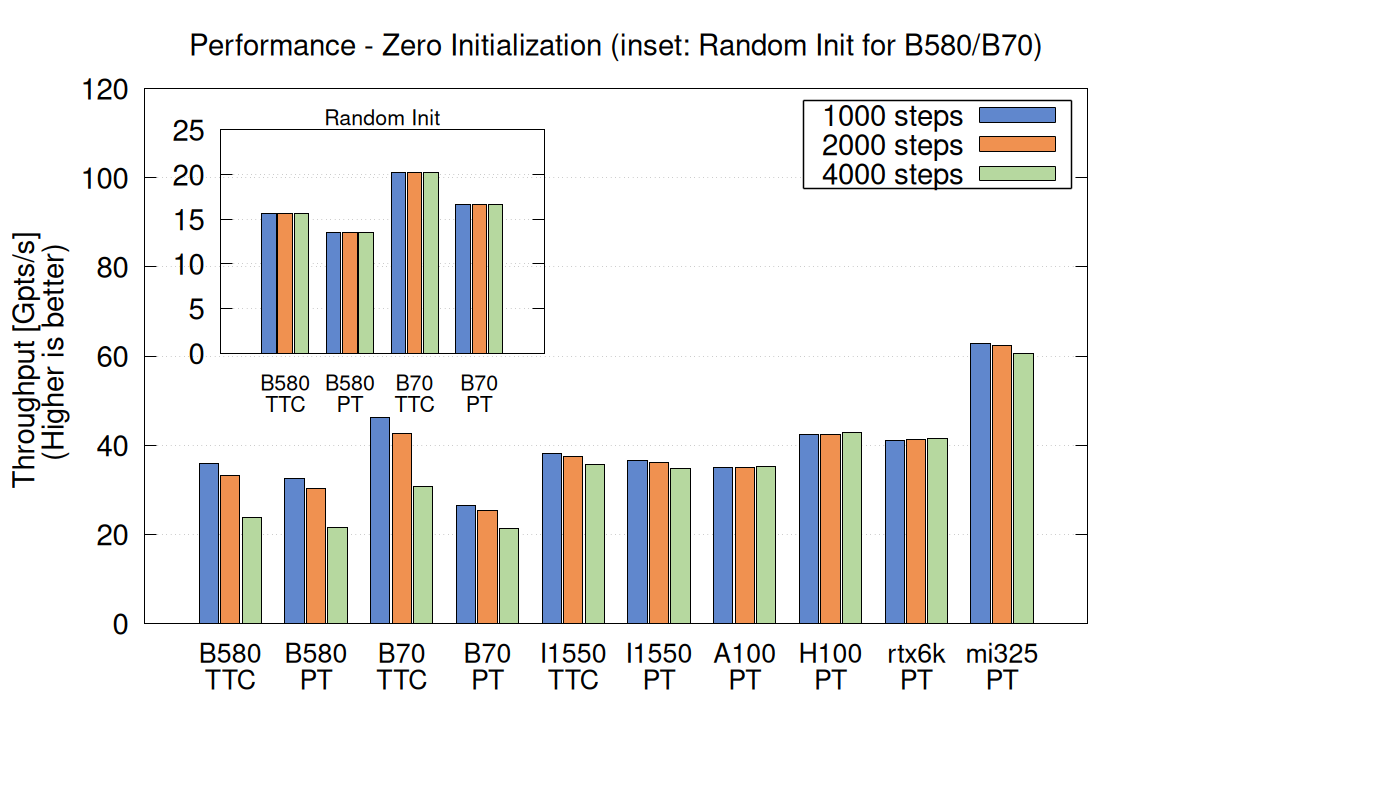}
 \caption{Throughput Gpts/s with zero-initialization
          for 1\,000, 2\,000, and 4\,000 time steps. 
          For a better comparison, the B580/B70 results 
          are replotted in the inset with random initialization, due to strong effect of the compression for zero initialization on the B580/B70.
}
 \label{fig:noinit}
\end{figure}

For the final result section, we benchmark all the architectures using
our PT/Triton implementation, which is portable across vendors.
Throughput results Gpts/s are plotted in Figure~\ref{fig:noinit}
and summarized in Table~\ref{tab:gpu_throughput}.

\begin{table}[t]
\centering
\caption{Throughput comparison Gpts/s across accelerators for 1,000, 2,000, and 4,000 steps. Memory compression is disabled on the Intel GPU MAX 1550 (GM1550). NVIDIA and AMD are insensitive to memory compression, therefore only one result is provided.}
\label{tab:gpu_throughput}
\footnotesize
\begin{tabular}{l ccc ccc}
\multirow{2}{*}{\textbf{Hardware}} & \multicolumn{3}{c }{\textbf{Zero-Initialization}} & \multicolumn{3}{c}{\textbf{Random Initialization}} \\
 & 1k & 2k & 4k & 1k & 2k & 4k \\
\midrule
B70 (TinyTC)     & 46.2 & 42.6 & 30.7 & 20.2 &  20.2 & 20.2 \\
B70 (PT/Triton)  & 26.4 & 25.3 & 21.2 & 16.6 & 16.6 & 16.6 \\
B580 (TinyTC)     & 35.8 & 33.1 & 23.8 & 15.6 & 15.6 & 15.6 \\
B580 (PT/Triton)  & 32.5 & 30.3 & 21.5 & 14.1 & 13.6 & 13.6 \\
GM1550 (TinyTC)     & 38.1 & 37.4 & 35.6 & 33.8 & 34.0 & 34.1 \\
GM1550 (PT/Triton)  & 36.6 & 36.1 & 34.7 & 32.9 & 33.1 & 33.1 \\
%\midrule
                   & \multicolumn{2}{c}{1k} & \multicolumn{2}{c}{2k} & \multicolumn{2}{c}{4k} \\
A100 (PT/Triton)   & \multicolumn{2}{c}{35.0} & \multicolumn{2}{c}{35.1} & \multicolumn{2}{c}{35.2} \\
H100 (PT/Triton)   & \multicolumn{2}{c}{42.4} & \multicolumn{2}{c}{42.5} & \multicolumn{2}{c}{42.9} \\
rtx6k (PT/Triton)  & \multicolumn{2}{c}{41.1} & \multicolumn{2}{c}{41.2} & \multicolumn{2}{c}{41.6} \\
mi325 (PT/Triton)  & \multicolumn{2}{c}{62.7} & \multicolumn{2}{c}{62.4} & \multicolumn{2}{c}{60.6} %\\ 
%\midrule
 %                   & \multicolumn{6}{c}{\textbf{Claude}} \\
  %                  & \multicolumn{2}{c}{1k} & \multicolumn{2}{c}{2k} & \multicolumn{2}{c}{4k} \\
%A100 (PT/Triton)    & \multicolumn{2}{c}{26.3} & \multicolumn{2}{c}{26.3} & \multicolumn{2}{c}{26.3} \\
%rtx6k (PT/Triton)   & \multicolumn{2}{c}{59.4} & \multicolumn{2}{c}{59.4} & \multicolumn{2}{c}{59.4} \\
%mi325 (PT/Triton)   & \multicolumn{2}{c}{104.1} & \multicolumn{2}{c}{103.1} & \multicolumn{2}{c}{100.0} \\
\end{tabular}
\end{table}

Comparing the B70 and B580 under zero initialization, both systems show the same
overall trend with TinyTC: as the simulation advances, the number of non-zero
elements grows and throughput gradually falls. Peak throughput reaches
46.2 and 35.8\,Gpts/s at 1\,000 steps for the B70 and B580, respectively,
which is a strong result for mid-range cards. The Python/Triton version follows
the same pattern and delivers comparable performance on the B580, although it is
about 8\% slower, most likely because of the 2D block-load strategy. The B70 is
more interesting: TinyTC behaves like the B580 but with a $\times 1.3$ speedup,
which closely matches the measured peak memory-bandwidth ratio between the two
cards ($540/400 \approx 1.35$).

PT/Triton produces a more mixed picture under zero initialization. On the B70,
performance falls below the B580 result, which is unexpected. Despite extensive
investigation, we could not identify the cause of this discrepancy.

The Intel GPU MAX 1550 (GM1550) behaves consistently with these observations.
As reported in~\cite{wassell2025optimized}, the sustained STREAM bandwidth of
the GM1550 is closer to $\sim\!1000\,\mathrm{GB/s}$ than to its theoretical
peak of $1600\,\mathrm{GB/s}$, while the B580 reaches about
$405\,\mathrm{GB/s}$ in practice. This gives a practical bandwidth ratio of
$1000/405 \approx 2.47$, which aligns well with our random-initialization
results: the GM1550/B580 throughput ratio is $32.9/13.5 \approx 2.44$ for
PT/Triton and $33.8/15.6 \approx 2.17$ for TinyTC.

In random initialization, the Intel results are consistent across devices.
The throughput ratios follow the measured bandwidth ratios within about 10\%:
GM1550/B70/B580 give relative ratios of $1.0/1.9/2.2$, respectively. The
B70 PT/Triton result is slightly lower than expected, but remains consistent
and is still clearly better than in the zero-initialization case.

The second part concerns NVIDIA and AMD hardware, tested with Python/Triton only.
The interpretation here is simpler: since these platforms do not implement memory compression,
zero-initialization and random-initialization yield identical results, 
and only one value is reported per configuration.

We compare PT/Triton results exclusively, as TTC is limited to Intel GPUs. The A100, H100, RTX 6000,
and MI325X execute the same Python/Triton code as the Intel hardware; however, 
performance is limited by the quality of the Triton backend on each platform 
and the available memory bandwidth, and falls short of what bandwidth ratios alone would suggest.
Taking the B580 as the reference point at 400 GB/s (measured bandwidth), the bandwidth ratios with respect to the
A100, H100, RTX 6000, and MI325X are respectively (4.5/4.8/3.7/11.3).

The performance ratios of the NVIDIA and AMD platforms relative to the B580 are (1.1/1.3/1.3/1.9) for zero-initialization and (2.5/3.0/2.9/4.4) for random-initialization. Two observations follow:
\begin{itemize}
    
\item For zero-initialization, the gap between the B580 and server-class GPUs remains limited, 
largely because memory compression acts as a hardware-level accelerator on Intel platforms.

\item With random-initialization, a more realistic benchmark, the performance gap widens, 
yet remains well below what raw bandwidth ratios would predict.
%This suggests we have reached the limits of a generic algorithm applicable
%across all platforms. Our PT/Triton implementation has been highly beneficial 
%for Intel GPUs but remains only partially satisfying for AMD and NVIDIA hardware.
\end{itemize}

The main explanations for the lack of performance (at least a 2X speedup is missing) 
on AMD/NVIDIA flagship GPUs may be rooted in
the difference in architectural design compared with Intel GPUs. Our implementation of the 
Micikevicius algorithm relies primarily on the hardware-managed cache hierarchy, allowing 
the compiler to optimize data locality automatically (whereas the original paper achieves 
this manually using shared memory). This strategy appears well suited to Intel GPUs, whose 
X\textsuperscript{e}-core has significantly more L1 cache associated with it compared to the L1 cache per 
CUDA core in NVIDIA's SM design. For NVIDIA and AMD GPUs, this automatic cache optimization 
approach may be suboptimal. Instead, a more fine-grained management of data movement in 
particular through the explicit use of shared memory (or LDS on AMD) would likely improve 
performance. Additionally, leveraging NVIDIA-specific mechanisms such as the Tensor Memory 
Accelerator (TMA) would further enhance performance on NVIDIA platforms.

This hypothesis warrants further investigation to confirm the architectural impact on performance
and to validate the proposed optimizations for each platform.

In conclusion, every platform requires architecture-specific adaptations to achieve peak performance. In 
this sense, we are approaching the limits of truly generic kernels; much like OpenMP, where 
portable code often sacrifices performance for generality. Achieving peak performance on 
dedicated hardware ultimately requires platform-specific optimizations.

%In conclusion, beyond raw FLOPS, the B580/B770 achieve competitive performance with generic 
%code. Although results are  below those of current flagship GPUs, the performance-
%per-dollar ratio is notably more favorable.

%\textcolor{red}{
%The third part concerns kernels generated with the help of claude and was evaluated
%only on the A100, RTX\,6000, and MI325X. The results outperformed on RTX\,6000 and MI325X
%the hand made solution but not on the A100. Overall, the results are
%encouraging, especially on the %RTX\,6000, which outperforms the %H100
%despite having lower memory bandwidth. The results are already respectable and
%could likely be improved further, particularly on the MI325X, where the total
%memory bandwidth is very high and still not fully exploited.}

%\begin{figure}
% \centering
% \includegraphics[scale=0.22]{plot/init.png}
% \caption{Throughput (Gpts/s) with random field initialization
%          for 1\,000, 2\,000, and 4\,000 time steps.}
% \label{fig:init}
%\end{figure}

%\subsection{Roofline Analysis}

\subsection{Programming model}

From a software-engineering perspective, model selection should be
driven by developer experience, target hardware, and legacy-code
constraints rather than by peak performance alone.

OpenMP offload remains the most practical, lowest-effort path for porting
an existing CPU stencil code to GPUs, enabling the same codebase to execute
on both CPUs and GPUs with minimal code modifications.
restructuring but delivering the lowest throughput in our study.

SYCL provides a portable C++ path with explicit control over memory
and execution, and in our results it offers clearly better performance
than OpenMP while remaining less intrusive than a full DSL rewrite.

PT/Triton offers a strong productivity--performance compromise: its
Python-based tile abstraction is easier to adopt, portable across
vendors, and substantially faster than directive-only offload.

TinyTC like any low level DSL provides the highest performance on Intel hardware by exposing
fine-grained control over tile shape and 2D block-load behavior, but
it is also the most demanding programming model among the four
implementations evaluated.

Objectively, when an application is already written in SYCL, TinyTC is
an ideal performance booster (like CUDA Tile IR for NVIDIA): it can be applied selectively to the
critical kernels while keeping the existing SYCL host framework.

In practice, a pragmatic strategy is to avoid rewriting an entire
application in a new language and instead focus optimization effort on
the dominant bottleneck kernels. When those kernels control end-to-end
runtime, investing in a lower-level model such as TinyTC is justified;
otherwise, Triton or OpenMP often provides a better overall
time-to-solution. In addition, OpenMP, SYCL, and TinyTC can be mixed
within the same code base; this interoperability was used directly in
this project.
%\textcolor{red}{
%Finally, Claude is the topic of the moment. With minimal effort, we already 
%achieve competitive performance. 
%Since the Claude training corpus is predominantly Python-based, it is a natural 
%candidate for generating Python code; extensions to C++ are also reasonable. 
%However, exotic domain-specific languages such as TinyTC are poorly represented 
%in training data, making Claude-based generation largely ineffective for such targets.
%Moreover, the Claude tuning approach remains essentially brute-force: it does not 
%account for the specificities of new hardware architectures, which are also 
%unlikely to be well represented in the training corpus.
%In conclusion, when properly guided, Claude is certainly full of promise.}

% ---------------------------------------------------------------
\section{Conclusion}
\label{sec:conclusion}
This paper demonstrates that tensor compilers are practical tools for
high-order HPC stencils, beyond their original AI focus.
On Intel hardware, a specialized approach such as TinyTC delivers the
best performance by exposing low-level control over tiling and memory
operations.
%At the same time, a generalist approach based on mainstream AI
%frameworks (PyTorch + Triton) combined with Claude provides an attractive
%productivity--portability trade-off while maintaining competitive
%throughput across vendors

The main takeaway is that these two paths are complementary rather than
exclusive: specialization maximizes peak performance, while generalist
frameworks maximize portability and development velocity.

Future work will extend full memory-traffic analysis beyond steady-state
inner kernels, optimize boundary kernels, investigate the triton 
performance following the vendors architecture and explore tighter
integration of Triton into production code.

\section*{Acknowledgment}
%
% uncomment for camera ready
%
%The writing of this manuscript was assisted by a large language model,
%which drafted paragraphs/figures/tables based on detailed instructions and
%data/technical content provided by the author.
%All experimental data, analysis, and conclusions are the author's own work.
%The authors are grateful to Adam Dziekonski (Intel) 
%for his constructive discussions on the Intel X\textsuperscript{e}\,core2 architecture.

% ---------------------------------------------------------------
\bibliographystyle{IEEEtran}
\bibliography{references}

\end{document}